# Debye temperature of nanocrystalline Fe-Cr alloys obtained by mechanical alloying

S. M. Dubiel[1*], B. F. O. Costa[2], J. Cieslak[1] and A. C. Batista[2]

[1]AGH University of Science and Technology, Faculty of Physics and Applied Computer Science, PL-30-059 Krakow, Poland

[2]CFisUC, Physics Department, University of Coimbra, P-3004-516 Coimbra, Portugal

Abstract

A series on nanocrystalline $Fe_{100-x}Cr_x$ alloys prepared by mechanical alloying was investigated with X-ray diffraction (XRD), scanning electron microscopy (SEM) and Mössbauer spectroscopy (MS) techniques. XRD and SEM were used to structurally characterize the samples whereas MS permitted phase analysis as well as determination of the Debye temperature, $\theta_D$. Concerning the latter, an enhancement relative to bulk $\theta_D$-values was revealed in the range of $\sim 40 \leq x \leq \sim 50$. In a sample of $Fe_{55.5}Cr_{44.5}$ two phases were detected viz. (1) crystalline and magnetic with $\theta_D=572$ (56) K and (2) amorphous and paramagnetic with $\theta_D=405$ (26) K.

[*]Corresponding author: Stanislaw.Dubiel@fis.agh.edu.pl

## 1. Introduction

A great interest in nanometer-sized materials (NMs) has been observed in recent years. It stems both from their scientific and industrial importance [1]. As a rule, physical properties of NMs are different than those of their bulk counterparts: their values can be both enhanced as well as diminished. Concerning the Debye temperature, $\Theta_D$, a subject of the present study, it is regarded as a very important quantity to characterize materials properties such as lattice vibrations of atoms and phase transitions. There is a great body of experimental evidence that free and loosely embedded nanometer-sized particles show a decrease of $\Theta_D$ e. g. Fe [2-4] and Ag, Al and Au [5], while for the well-embedded, interacting or substrate-deposited metallic particles both increase or decrease of $\Theta_D$ is possible depending on the matrix or substrate. If the matrix is harder than the particles, then $\Theta_D$ of the latter is enhanced e. g. Sn in Si matrix [6], and vice versa e. g. Co in Ag matrix [7]. For NMs obtained by mechanical alloying (MA) the relationship between $\Theta_D$ and the size of grains, <d>, is ambiguous as according to some results $\Theta_D$ is increasing, e. g. [8], while according to other ones it is decreasing, e. g. [9], with decreasing grain size. The discrepancy may be, at least partly, due to different methods applied for determining $\Theta_D$, because they are sensitive to different lattice vibrations frequencies, and/or they probe different structural properties of nanoparticles/grains. In fact, the Debye temperature of nano crystalline titania determined with the X-ray diffraction (XRD) technique was found to increase gradually with decreasing size of grains, while its values obtained from differential scanning calorimetric measurements exhibited a reversed behavior [10]. A recent study of Fe-Cr alloys obtained by MA gave evidence that the $\Theta_D$-values determined from the intensities of the XRD-peaks were higher than their counterparts found for the bulk Fe-Cr alloys [11].

Also the size and shape of n-clusters plays a role as far as the values of $\Theta_D$ are concerned [12]. Results depicting the Debye temperature obtained with the Mössbauer spectroscopy (MS) on a series of nanometer-sized $Fe_{100-x}Cr_x$ alloys (0 ≤ x ≤ 84), hereafter called n- $Fe_{100-x}Cr_x$, obtained by mechanical alloying (MA) are reported in this paper. Although n-FeCr alloys prepared by MA were often studied by MS e. g. [13-22], there is not, to our best knowledge, such study on the Debye temperature known in the available literature.

## 2. Experimental

The n- $Fe_{100-x}Cr_x$ alloys, subject of this investigation, were prepared by mechanical alloying (MA) of Fe (99.0+%, grain size 60 μm) and Cr (99.0+%, grain size 45 μm) elemental powders in Ar atmosphere using a planetary mill (Fritsch-P6) at a disc rotation speed of 500 rpm. The mill was equipped with hardened steel vials and seven balls. The powder to balls mass ratio was 1:20. The total milling time was 16h at maximum, but it was interrupted every hour for 15 min. The total milling time was not the same for all samples in order to obtain samples with different crystallite sizes, but the time was always been sufficient to form an alloy, as known from our previous studies in mechanosynthesis of Fe-Cr alloys [22-25]. The chemical composition analysis of the samples was performed by electron probe microanalysis (EPMA) with a Camebax SX 50 system – see Table 1. The inhomogeneity in the composition was less than 0.2at%. The samples were investigated by scanning electron microscopy (SEM), X-ray diffraction (XRD) and Mössbauer spectroscopy (MS). The electron microscope used was a FEI Quanta 400FEG. SEM pictures, examples of which are shown in Fig. 1, gave evidence that the n-Fe-Cr alloys had a form of agglomerates of micrometric sizes.

XRD was performed on the samples at room temperature (RT) using Cu Kα radiation (λ = 0.154184 nm). The mean crystallite size, <d>, and microstrain, ε, of the samples were obtained from the widths of the XRD peaks using the Williamson-Hall method [26]. XRD diffraction patterns are partly shown in Fig. 2. Lattice constant, *a*, average size of crystallites, <d>, and microstrain, ε, are displayed in Table 1.

$^{57}$Fe-site Mössbauer spectra were recorded in the temperature interval of 80-300 K and in a transmission geometry using a standard spectrometer with a sinusoidal drive. Gamma rays of 14.4

keV energy were supplied by a Co/Rh source having a nominal activity of 50 mCi. The Mössbauer spectra were analyzed by means of a standard hyperfine field distribution (HFD) method [27] assuming a linear correlation between the hyperfine field and the isomer shift [28]. The analysis yielded values of the center shift, CS, as well as HFD histograms. From the temperature dependence of the former, the Debye temperature, $\theta_D$, was determined as explained in section 3.2.

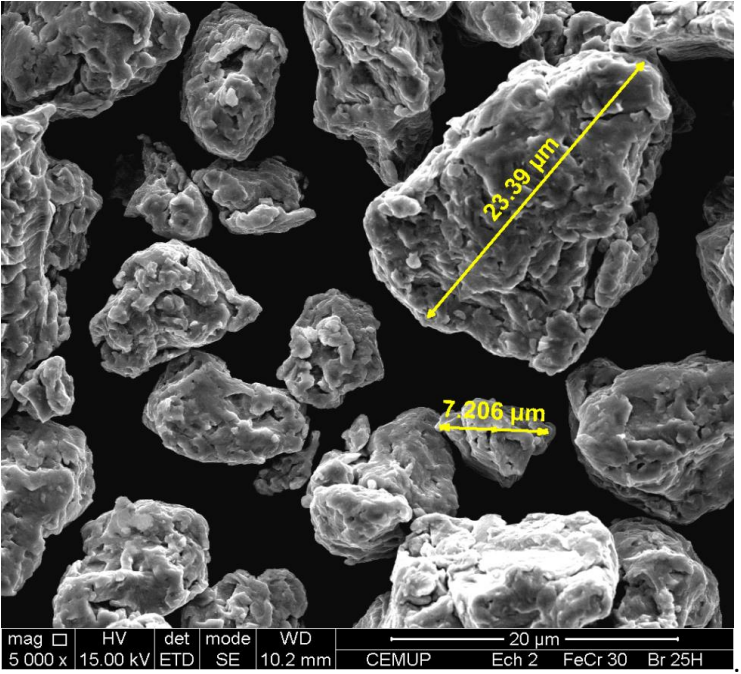

Figure 1 (a). SEM picture as revealed for the n-$Fe_{71.1}Cr_{28.9}$ alloy.

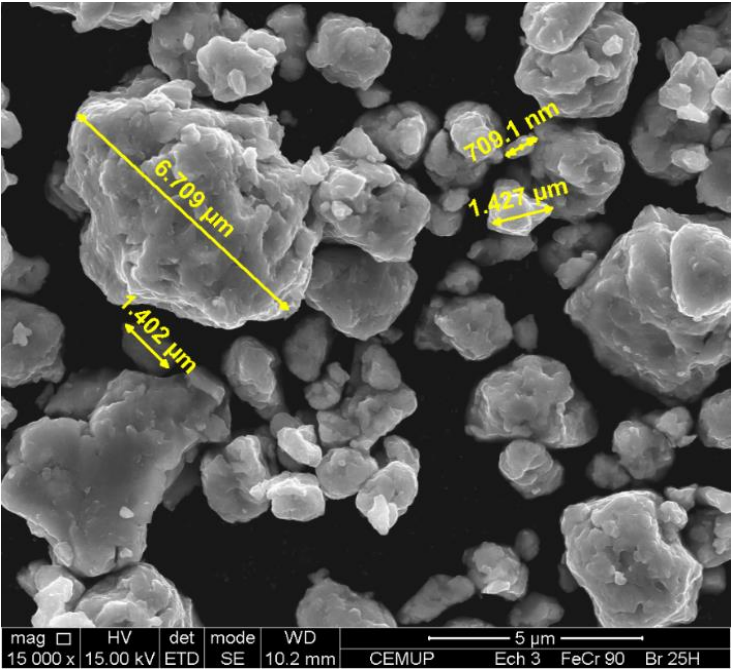

Figure 1(b). SEM picture as revealed for the n-$Fe_{16.2}Cr_{83.8}$ alloy.

Table 1
Parameters relevant to the investigated magnetic phase of n-Fe$_{100-x}$Cr$_x$ alloys: <d> - mean crystallite size, *a* – lattice constant, ε - microstrain, Θ$_D$ – Debye temperature.

| x[at%] | <d> [nm] | *a* [nm]     | ε[%]  | Θ$_D$[K]  |
|--------|----------|--------------|-------|-----------|
| 0      | 7.1(1)   | 0.28679(6)   | 1.40  | 504(13)   |
| 0      | 19.8(1)  | 0.28658(2)   | 0.76  | 505(25)   |
| 16.8   | 5.1(1)   | 0.28756(2)   | 1.24  | 446(42)   |
| 30.0   | 10.0(1)  | 0.28722(3)   | 1.03  | 467(37)   |
| 41.8   | 8.0(1)   | 0.28736(13)  | 0.975 | 509(49)   |
| 42.8   | 8.2(1)   | 0.28772(2)   | 0.91  | 509(32)   |
| 43.0   | 6.5(1)   | 0.28784(12)  | 1.31  | 552(52)   |
| 44.5   | 6.5(1)   | 0.28766(4)   | 1.1   | 484(22)   |
| 44.5   | 10.5(1)  | 0.28737(5)   | 0.61  | 578(79)   |
| 45.5   | 6.5(1)   | 0.28810(22)  | 1.32  | 541(42)   |
| 45.5   | 9.2(1)   | 0.28805(10)  | 0.88  | 572(26)   |
| 83.8   | 4.6(1)   | 0.28873(14)  | 1.35  | 446(16)   |
| 83.8   | 10.5(1)  | 0.28807(4)   | 1.28  | 452(12)   |

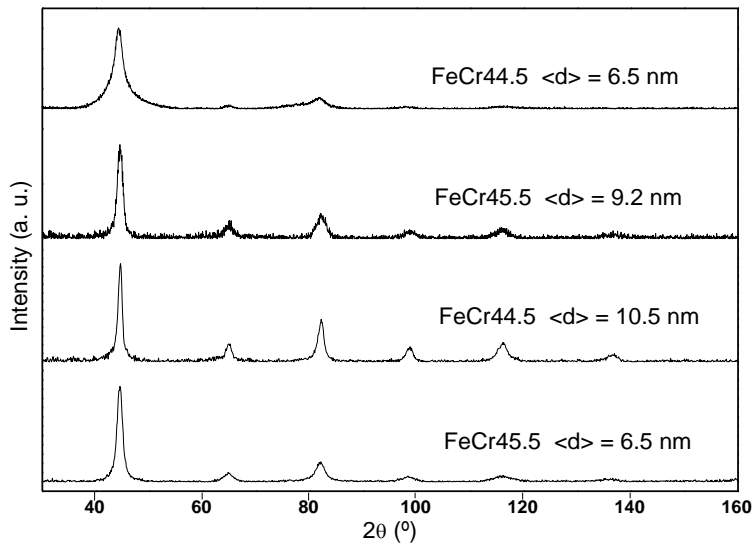

Figure 2. Examples of the XRD patterns recorded on n-Fe$_{55.5}$Cr$_{44.5}$ and n-Fe$_{55.5}$Cr$_{44.5}$ samples with different mean crystallite sizes, <d>, as indicated.

## 3. Results and discussion

### 3.1 XRD

Figure 2 shows examples of XRD patterns of n-Fe$_{55.5}$Cr$_{45.5}$ and n-Fe$_{56.5}$Cr$_{44.5}$ samples with different mean crystallite sizes, <d>. The positions of the peaks are in all cases (even for samples whose patterns are not shown) characteristic of a bcc phase but the peaks are broadened because of the nanocrystalline nature of the samples, yielding the values of mean crystallite sizes, <d> and microstrains, ε, as explained before. However, some patterns show an amorphous phase. Figures 3a and 3b present the strongest peak of the patterns from Fig. 2 for the n-Fe$_{56.5}$Cr$_{44.5}$ samples with <d>= 6.5 nm and <d>=10.5 nm, respectively, analyzed with two Lorentzian-shaped lines. The broadest line has about 65% of the peak area in the case of the sample with <d>= 6.5 nm, and about 30% of the peak area, in the case of the sample with <d>=10.5 nm. These broad lines can be ascribed to an amorphous phase. Amorphization is an effect known to occur during mechanical alloying. For Fe-Cr alloys, several studies conducted by ball milling showed their amorphization e.g. [13-21]. In particular, in equiatomic or near-equiatomic FeCr alloys milled in vacuum or argon atmosphere, this effect was observed both in alpha and sigma phases, e.g. [29-32].

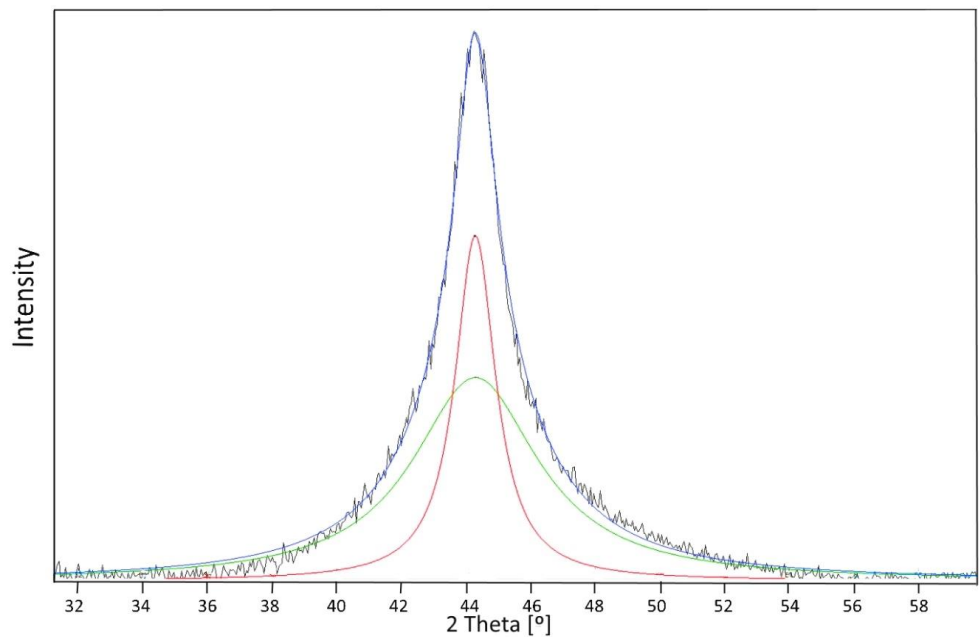

Figure 3 (a). The strongest peak of the XRD pattern shown in figure 2 for the n-$Fe_{55.5}Cr_{44.5}$ sample with <d>= 6.5 nm, analyzed with two Lorentzian-shaped lines.

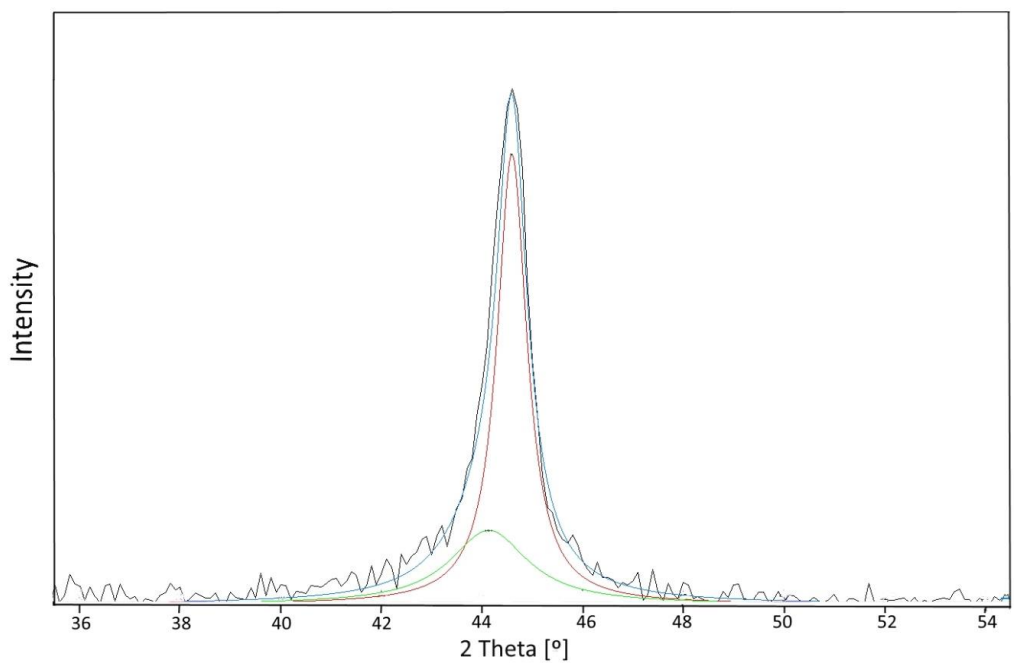

Figure 3 (b). The strongest peak of the XRD pattern shown in figure 2 for the n-$Fe_{55.5}Cr_{44.5}$ sample with <d>= 10.5 nm, analyzed with two Lorentzian-shaped lines corresponding to the crystalline and amorphous phases, respectively.

## 3.2. Mössbauer Spectroscopy

Examples of the Mössbauer spectra recorded at RT are shown in the left panel of Fig. 4a (single-phased samples), and in the left panel of Fig. 4b (two-phased samples).

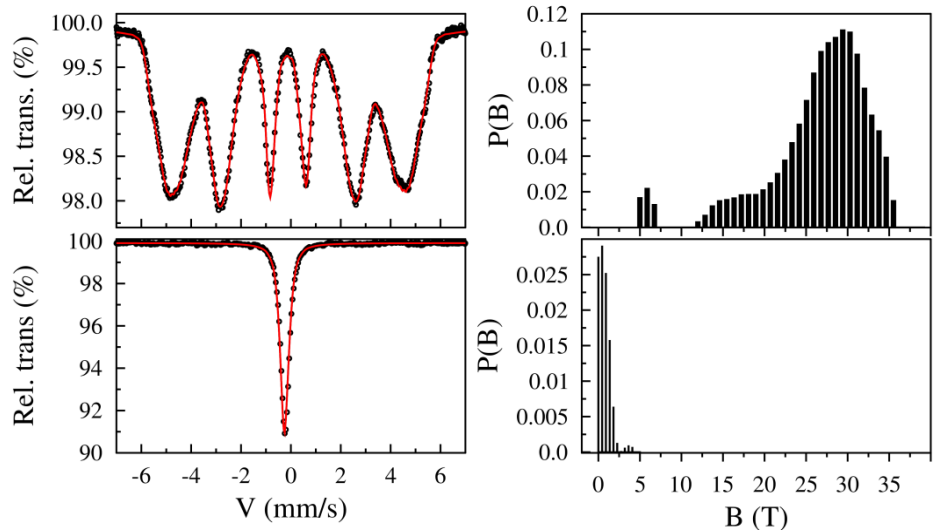

Figure 4a. (left panel) $^{57}$Fe Mössbauer spectra recorded on the n-Fe$_{83.2}$Cr$_{16.8}$ sample with <d>=5.1 nm (top), and that on the n-Fe$_{16.2}$Cr$_{83.8}$ sample with <d>=4.6 nm (bottom). The corresponding HFD histograms are illustrated in the right panel.

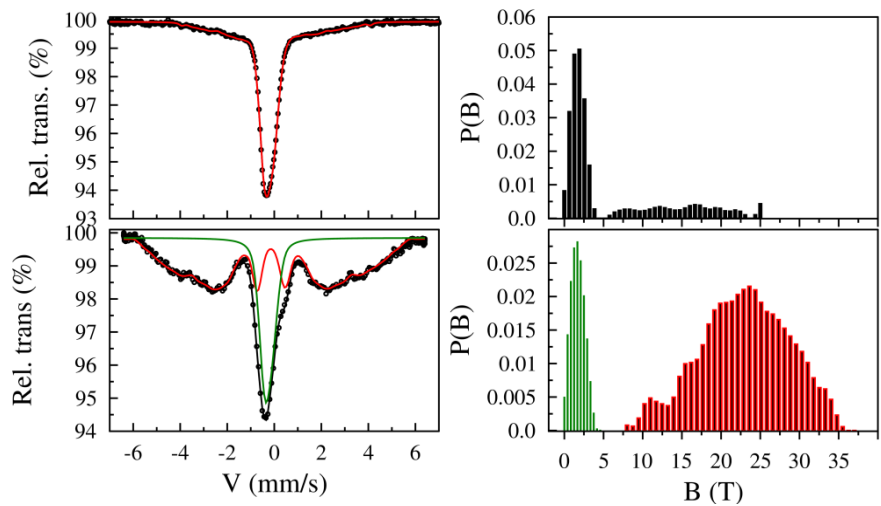

Figure 4b. (left panel) $^{57}$Fe Mössbauer spectra recorded on the n-Fe$_{55.5}$Cr$_{44.5}$ sample with <d>=6.5 nm (top) and <d>=10.5 nm (bottom). The corresponding HFD histograms are shown in the right panel.

The spectra showing only one (magnetic or paramagnetic) phase, like the ones shown in Fig. 4a, were analyzed in terms of one HFD, while those in which two phases were observed, like the ones presented in Fig. 4b, were analyzed with two HFDs, one for the paramagnetic phase and another for the magnetic phase HFD-histograms were obtained as outputs of such fits, and their examples are shown in Fig. 4a and 4b (right panels).

The Debye temperature, $\Theta_D$, can be determined from the Mössbauer spectra in two alternative ways viz. either from a temperature dependence of the center shift, *CS*, or from that of the recoil-free factor (spectral area). Here we applied the former way i.e. $\Theta_D$ was determined using the following equation:

(1)
$$CS(T) = IS(0) - \frac{3kT}{2mc}\left[\frac{3\Theta_D}{8T} + \left(\frac{T}{\Theta_D}\right)^3 \int_0^{\Theta_D/T} \frac{x^3}{e^x - 1}dx\right]$$

Where *IS* stays for the isomer shift (temperature independent), *k* is the Boltzmann constant, *m* is a mass of $^{57}$Fe atoms.

An example of the *CS(T)*-dependence obtained for the two-phased n-Fe$_{55.5}$Cr$_{45.5}$ sample with <d>=10.5 nm can be seen in Fig. 5 together with the best fits to the experimental data in terms of eq. (1). It is clear that the two phases have different values of *CS*, hence Fe-site charge-densities. They have also different values of $\Theta_D$ viz. 578(79) K for the magnetic phase (see table 1), and 405(30) K for the non-magnetic one. Different $\Theta_D$-values mean different lattice dynamics or degree of lattice stiffness. Possible reasons underlying this difference are discussed below.

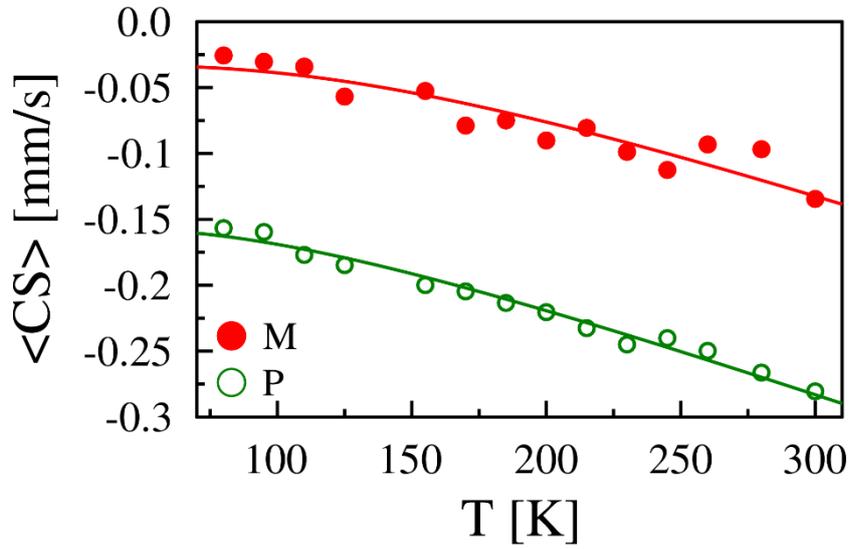

Figure 5. The average center shift of the magnetic (M) and paramagnetic (P) sub-spectra, <CS>, versus temperature, *T*, as found for the n-Fe$_{55.5}$Cr$_{44.5}$ sample with <d>=10.5 nm. The solid lines represent the best fits to the data in terms of eq. (1).

The values of $\Theta_D$ obtained from these fits for the all studied n-Fe-Cr alloys are displayed in Fig. 6. The corresponding ones found previously for the bulk Fe-Cr alloys using the same method [33] as well as those determined with the XRD for MA Fe-Cr alloys [11] are added for the sake of comparison. It is evident that the values of $\Theta_D$ as presently determined for pure iron and for the n-Fe-Cr alloys with the composition close to the near-equiatomic one are meaningfully higher than those revealed for the bulk alloys, while for the other composition they are in line with those found for the bulk alloys. It is not clear why this difference occurs, but also the XRD-data reported in Ref.11 exhibited enhanced values for Cr compositions above ~40%. Noteworthy, we did not find any difference in $\Theta_D$ between nano-crystalline and bulk alloys of σ-Fe-Cr alloys [34]. However, the observed enhancement of $\Theta_D$ agrees both with experimental observation according to which various physical properties determined for nanometer-sized alloys obtained by MA had enhanced values relative to the corresponding ones found for the bulk counterparts e. g. [35] as well as with model calculations e. g. [36]. According to the latter the enhancement occurs when nanoparticles or nanograins interact with each other e.g. they form aggregates (as in the present case) or with a substrate onto which they had been deposited or with a matrix into which they had been embedded. However, it is not clear why the enhancement is observed only for the samples with the near-equiatomic compositions whereas the agglomerates were formed for all studied samples.

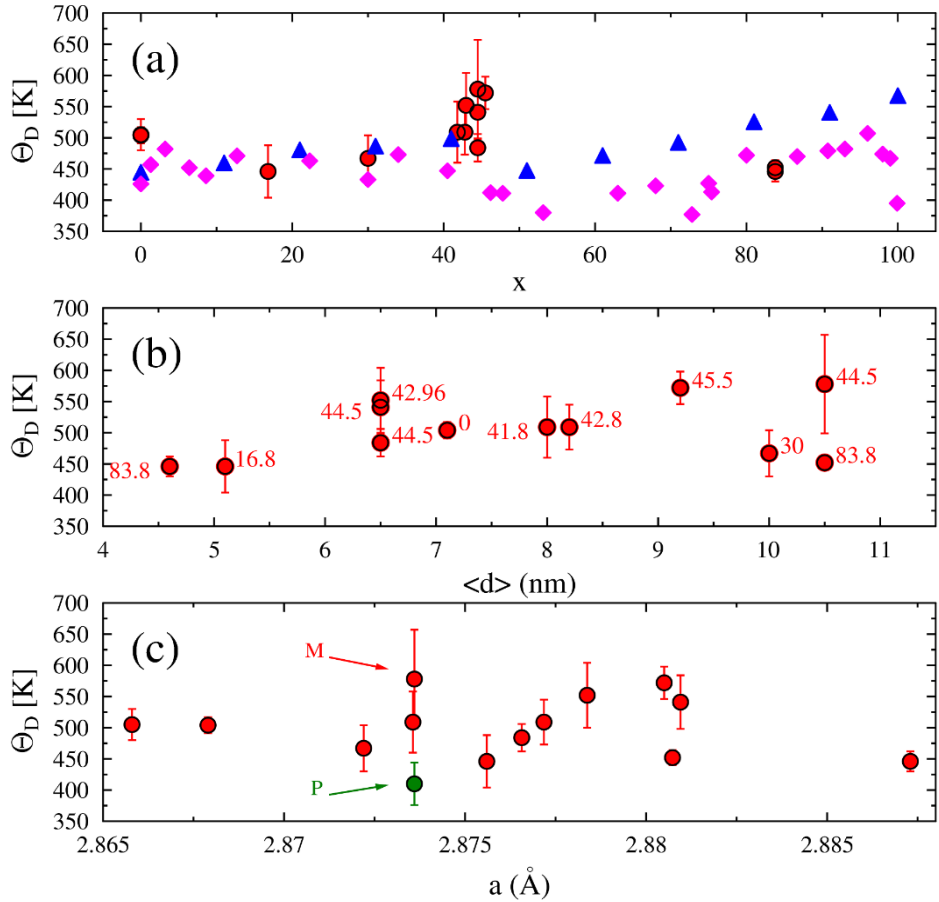

Figure 6. The Debye temperature, $\Theta_D$, vs. (a) $x$ for the presently studied n-$Fe_{100-x}Cr_x$ samples (circles) and that found previously for the bulk Fe-Cr [32/33] (diamonds) and MA Fe-Cr alloys [11] (triangles), (b) <$d$> and (c) $a$ for the n-$Fe_{100-x}Cr_x$ alloys. In the latter case the values of $\Theta_D$ derived from the paramagnetic (P) and the magnetic (M) sub spectra recorded on the n-$Fe_{55.5}Cr_{44.5}$ sample with <$d$>=10.5 nm are marked.

For non-interacting nanoparticles $\Theta_D$ decreases, as a rule, with decreasing <$d$>. Here, as evidenced in Fig. 6b, it is not the case: for the same or similar compositions, $\Theta_D$ does not depend on the size of grains. This feature is especially well pronounced for $x$=83.8 for which <$d$> changes between 4.6 and 10.5 nm, a range in which the grain size dependence of $\Theta_D$ for non-interacting n-particles is the strongest [2-4], while in this study $\Theta_D$ is the same for both sizes. For the samples with $41.8 \leq x \leq 45.5$, an enhancement of the $\Theta_D$ – values relative to their bulk counterparts is observed. The most significant difference in $\Theta_D$ exists between the non-magnetic (P) and magnetic (M) phases revealed in the n-$Fe_{55.5}Cr_{44.5}$ sample. It may be understood in terms of the underlying difference in a degree of crystallinity of the two phases, which in turn, can be associated with a well-crystallized grain cores (M), and a highly-disordered or even amorphous phase (P). Such interpretation is supported by the XRD-patterns, as discussed in section 3.1. Further arguments in favor of the existence of the amorphous phase follows from an alternative analysis of the Mössbauer spectra recorded for the n-$Fe_{55.5}Cr_{44.5}$ samples. The non-magnetic sub spectrum (P) was analyzed in terms of the quadrupole splitting distribution, QSD, while the magnetic one (M) in terms of the HFD. The obtained HFD and QSD histograms for both samples i.e. that with <$d$>=6.5 nm and the one with <$d$>=10.5 nm are displayed in Fig. 8, while obtained spectral parameters are shown in Table 2.

Table 2. Spectral parameters as found from the analysis of the Mössbauer spectra of the n-Fe$_{55.5}$Cr$_{45.5}$ samples in terms of the HFD method (M subspectrum) and in terms of the QSD method (P subspectrum). The average values of the hyperfine field, <B>, and those of the quadrupole splitting, <QS>, were calculated by integrating the HFD and QSD histograms, respectively, shown in Fig. 8. The relative spectral areas of M and P sub spectra are marked by A, and the average isomer shift by <IS> (relative to the α-Fe).

| Phase | Spectral parameter | <d>=6.5 nm | <d>=10.5nm |
|---|---|---|---|
| M | <B> [T] | 13.9 | 22.9 |
|   | <IS> [mm/s] | 0.03 | -0.023 |
|   | A [%] | 31 | 72 |
| P | <QS> [mm/s] | 0.38 | 0.39 |
|   | <IS> [mm/s] | -0.121 | -0.160 |
|   | A [%] | 69 | 28 |

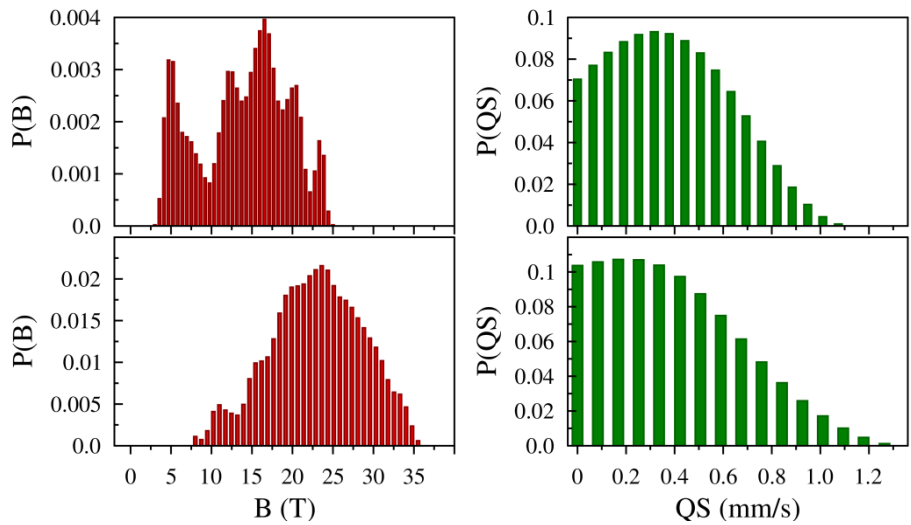

Figure 7. HFD histograms (left panel) and QSD histograms (right panel) obtained by analyzing the magnetic sub spectra of the Mössbauer spectra recorded on the n-Fe$_{55.5}$Cr$_{45.5}$ samples in terms of the HFD method and the non-magnetic ones in terms of the QSD method. Upper diagrams are for <d>=6.5 nm while the lower ones are for <d>=10.5 nm.

It is observed that the shape and the range of QS-values seen in the presently determined QSD histograms agree well with those previously reported [15, 16, 21] for amorphous Fe-Cr alloys. From the data displayed in Table 2 one can see that the abundances of the M-phases agree fairly well with the intensities of the narrow diffraction peak, while those of the P-phases are in line with the intensities of the broad XRD peak. Consequently, we can ascribe the magnetic sub spectrum to the crystalline phase, while the paramagnetic one to the amorphous phase. The values of the average isomer shift, <IS>, ascribed to the P-phase have significantly more negative values than the corresponding ones for the M-phase i.e. the charge-density of s-like electrons at Fe-nuclei is higher in the amorphous (P) than in the crystalline (M) phase. This effect may be related to a more densely packed amorphous structure. Moreover, a higher amount of the P-phase found in the <d>=6.5 nm sample (65-69%) than that in the <d>=10.5 nm sample (28-30%) supports the outlined interpretation as the amount of the amorphous phase (P-phase) should be proportional to the milling time.

## 4. Summary

The results obtained in this study can be summarized as follows:

1. Nano crystalline $Fe_{100-x}Cr_x$ alloys with $0 \leq x \leq \sim 84$ and the average grain size between 4.6 and 19.8 nm were obtained by mechanical alloying.

2. The nanocrystals exist in form of conglomerates with diameter of up to $\sim 25 \mu m$.

3. Some of the samples were found to be two-phased i.e. crystalline and amorphous.

4. Debye temperature, $\Theta_D$, was determined from the temperature dependence of the center shift.

5. An enhancement of $\Theta_D$ relative to bulk counterparts was found for the samples with $41.8 \leq x \leq 45.5$.

6. The Debye temperature of the crystalline and magnetic phase is significantly higher than the one of the amorphous and paramagnetic phase.

**Acknowledgements**

This work was supported by funds from FEDER (Programa Operacional Factores de Competitividade COMPETE) and from FCT - Fundação para a Ciência e Tecnologia - under the project PEst-C/FIC/UI0036/2011. We also acknowledge a bilateral Polish-Portuguese collaboration between the Polish Ministry of Science and Higher Education and FCT in the years 2007-2010.